\documentclass[times]{tum-article}

\usepackage{cite}

\graphicspath{{plots/}}

\title{Analyzing the Positivity Preservation of Numerical Methods for the Liouville-von Neumann Equation}
\author{Michael Riesch\orcid{0000-0002-4030-2818}\authormark{\Letter} and Christian Jirauschek\orcid{0000-0003-0785-5530}}
\affil{Department of Electrical and Computer Engineering, Technical University of Munich, Arcisstr. 21, 80333 Munich, Germany}
\email{michael.riesch@tum.de}
\date{Received: 01 August 2018 / Accepted: 03 April 2019 / Available online:
  08 April 2019
  \thanks{This is a post-peer-review, pre-copyedit version of an article
    published in Journal of Computational Physics. The final authenticated
    version is available online at:
    \url{http://dx.doi.org/10.1016/j.jcp.2019.04.006}}}

\DeclareMathOperator{\col}{col}

\begin{document}

\maketitle

\begin{abstract}
The density matrix is a widely used tool in quantum mechanics. In order to determine its evolution with respect to time, the Liouville-von Neumann equation must be solved. However, analytic solutions of this differential equation exist only for simple cases. Additionally, if the equation is coupled to Maxwell's equations to model light-matter interaction, the resulting equation set -- the Maxwell-Bloch or Maxwell-Liouville-von Neumann (MLN) equations -- becomes nonlinear. In these advanced cases, numerical methods are required. Since the density matrix has certain mathematical properties, the numerical methods applied should be designed to preserve those properties. We establish the criterion that only methods that have a completely positive trace preserving (CPTP) update map can be used in long-term simulations. Subsequently, we assess the three most widely used methods -- the matrix exponential (ME) method, the Runge-Kutta (RK) method, and the predictor-corrector (PC) approach -- whether they provide this feature, and demonstrate that only the update step of the matrix exponential method is a CPTP map.

\end{abstract}

\let\vec\relax

\section{Introduction}
\label{sec:introduction}

In quantum mechanics, the density matrix $\hat \rho$ is a widely used concept to describe an ensemble of quantum states. The Liouville-von Neumann equation
\begin{equation}
  \partial_t \hat \rho = \mathcal{L}\left(\hat\rho\right),
  \label{eq:lvn}
\end{equation}
where $\mathcal{L}$ is the linear Liouvillian superoperator, plays a crucial role in this context as it determines the temporal evolution of the density matrix~\cite{tang2005}. Equation~(\ref{eq:lvn}) may be combined with Maxwell's equations to form the (generally nonlinear) Maxwell-Liouville-von Neumann (MLN) equations, which are used to model light-matter interaction in systems where a quantum mechanical description of matter is required. In the scope of our research, the MLN equations describe the dynamics of quantum cascade lasers (QCLs)~\cite{jirauschek2014}. Mostly, the rotating wave approximation (RWA) is used in related work to simplify the MLN equations and solve them analytically. However, this approximation cannot be applied in the scope of our work since the electric field in QCLs may feature broad spectra and/or high peak intensities, which renders the RWA invalid~\cite{riesch2018atrasc1}.

For both the Liouville-von Neumann equation and the MLN equations analytic solutions can be derived only for very specific cases. Therefore, numerical methods are generally required to solve them. Several methods are presented in related literature and can be divided into three main categories. Ziolkowski, Slavcheva and coworkers used a Crank-Nicolson (CN) scheme in their work, where the implicit nature of the CN scheme was resolved with a predictor-corrector (PC) approach~\cite{ziolkowski1995, slavcheva2002}. In the work at hand, we will treat these two methods separately. The methods of the second category~\cite{hellsing1986, berman1992, bidegaray2003, saut2006, marskar2011} solve Eq.~(\ref{eq:lvn}) by calculating the matrix exponential $\exp(\mathcal{L}t)$. This category is referred to as ME methods in the following. Finally, several implementations of the Runge-Kutta (RK) method have been presented (see e.g.~\cite{garraway1994, sukharev2011, deinega2014, cartar2017}).

Naturally, the efficiency and the accuracy of the numerical methods are crucial. The density matrix is by definition Hermitian, positive semi-definite, and its trace equals 1~\cite{tang2005, kraus1983}. In order to guarantee realistic results, these properties must be preserved by the equations as well as by the numerical methods. Bid\'egaray et al.~\cite{bidegaray2001} analyzed the positivity preservation of the CN scheme and found that this method does not necessarily yield a positive semi-definite density matrix. In a subsequent publication, Songolo and Bid\'egaray stated that the Runge-Kutta method does not preserve the properties of the density matrix but no rigorous proof was given~\cite{songolo2018}. To the best of our knowledge, such a  rigorous analysis for the PC and RK methods has not been published. While the ME methods preserve the properties by definition in theory, the actual implementations must be analyzed since several approximation technique are applied in practice. As to the efficiency of the numerical methods, we recently performed a performance comparison of the PC, ME, and RK methods and found that both the Runge-Kutta and the predictor-corrector implementation outperformed the matrix exponential implementation~\cite{riesch2018oqel}. However, the correctness of both methods has not been verified yet.

In the work at hand, we first establish the criterion that the update of a numerical method from time step to time step must be a completely positive trace preserving (CPTP) map. If this criterion holds, the method is suitable for long-term simulations in which we are particularly interested. Subsequently, we analyze whether or not the different numerical methods (PC, ME, RK) fulfill this criterion. Then, by implementing a simulation setup from related literature, we confirm the results of the theoretical analysis in practice. Finally, we conclude with our findings and present an outlook on future work.

\section{Theoretical Background}

In this section we provide the foundations that are required in the following sections. First, we discuss the properties of the density matrix and their consequences for the Liouville-von Neumann equation. A description of different representations of the density matrix follows. A short summary of completely positive trace preserving (CPTP) maps concludes this section.

\subsection{Density matrix and Liouville-von Neumann equation}

The properties of the density matrix have already been discussed shortly in Section~\ref{sec:introduction}. The density matrix $\hat \rho$ is Hermitian, i.e., $\hat \rho = \hat \rho^\dagger$, where the dagger denotes the conjugate transpose. It is also positive semi-definite, which can be denoted as $\hat \rho \geq 0$. Finally, its trace $\trace\{ \hat \rho(t) \} = 1$ must remain constant.

Therefore, it becomes apparent that the Liouvillian superoperator $\mathcal{L}$ on the right-hand side of the Liouville-von Neumann equation cannot be chosen arbitrarily. For example, in order to fulfill the trace condition of the density matrix, the trace of the superoperator $\trace\{\mathcal{L}(\hat\rho)\} = 0$ must vanish. In the following, we consider the expansion of Eq.~(\ref{eq:lvn}) to the Lindblad form
\begin{equation}
\partial_t \hat \rho = \mathcal{L} \left( \hat \rho \right) = -\mathrm{i}\hslash^{-1} \left[ \hat H, \hat \rho \right] + \mathcal{G} \left( \hat \rho \right),
\label{eq:lvn2}
\end{equation}
which is guaranteed to preserve the properties of the density matrix~\cite{lindblad1976, gorini1976}. Here, $\hslash$ is the reduced Planck's constant, $[\cdot,\cdot]$ denotes the commutator, $\hat H$ is the Hamiltonian of the system, and $\mathcal{G}$ is the relaxation superoperator.

\subsection{Density matrix in different representations}

Normally, the density matrix of a system with $N$ energy levels is represented by a $N \times N$ matrix. As we shall see, it may be beneficial to switch to a vector representation. For example, the columns of the density matrix $\hat \rho$ can be stacked on top of each other into a vector $\vec \rho$ with $N^2$ complex elements. This column-major order is denoted as $\vec \rho = \col(\hat \rho)$ and called representation in Liouville space~\cite{marskar2011}. The Liouville-von Neumann equation then reads $\partial_t \vec \rho = L \vec \rho$, where $L$ is a $N^2 \times N^2$ matrix representing the Liouvillian $\mathcal{L}$.

\subsection{Completely positive trace preserving maps}

In general, it is not sufficient that the solutions of the Liouville-von Neumann equation are positive maps. Indeed, every solution must constitute a completely positive trace preserving (CPTP) map in order to preserve the density matrix properties~\cite{lindblad1976, lebellac2011}. A map $\mathcal{U}$ is CPTP if and only if a decomposition
\begin{equation}
  \mathcal{U}\left(\hat \rho\right) = \sum_{i=1}^{N^2} \hat V_i \hat \rho \hat V_i^\dagger
  \label{eq:cptp}
\end{equation}
exists~\cite{kraus1971, choi1975}, where $N$ accounts for the dimension of the Hilbert space (which corresponds to the number of considered energy levels), and the Kraus operators $V_i$ fulfill
\begin{equation}
  \sum_{i=1}^{N^2} \hat V_i^\dagger \hat V_i = \hat I,
  \label{eq:cptp2}
\end{equation}
where $\hat I$ is the $N \times N$ identity matrix. See e.g.~\cite{nielsen2010} for a detailed description of CPTP maps.

\section{Analysis of the Numerical Methods}
\label{sec:analysis}

Every numerical method performs an update of the density matrix $\hat \rho_n = \hat\rho \left( t_n \right)$ at every time step $t_n = n \Delta t$, where $\Delta t$ is the time step size. This update can be written as map $\hat \rho_{n + 1} = \mathcal{U}_n( \hat\rho_n )$. If this map of a certain numerical method is CPTP, the properties of the density matrix are guaranteed to be preserved over the complete simulation time span. Conversely, a numerical method that does not feature a CPTP update map may yield unrealistic results (for example, a violation of the trace condition as reported in~\cite{bidegaray2001}). The methods discussed in this section have been tested by their authors and successfully applied to the respective problems. Therefore, we assume that a parameter region exists (in particular with respect to the simulation end time) where all numerical methods produce reasonable results. However, we shall not discuss possible regions of validity but use the CPTP map criterion above to evaluate strictly whether the numerical methods return realistic results for all input parameters or not. Based on this evaluation, we shall continue to use the method(s) that passed for our long-term simulations (i.e., large simulation end times). In the following, we discuss three numerical methods and their variations: the matrix exponential (ME) approaches, the Runge-Kutta (RK) method, and the predictor-corrector (PC) technique.

\subsection{Matrix exponential approaches that solve the equation exactly for
a single time step}

The numerical methods of this group calculate the exponential $\exp(\mathcal{L}_n\Delta t)$ of the Liouvillian $\mathcal{L}_n$ in order to solve the differential equation exactly at every time step $n$. The form of this exponential (and therefore the form of the update map) is determined by the representation. For example, in Liouville space the update reads
\begin{equation}
  \vec\rho_{n+1} = \exp\left(L_n \Delta t\right) \vec\rho_n \coloneqq U_n \vec\rho_n,
  \label{eq:update-me}
\end{equation}
where $L_n$ and $U_n$ are the matrices that represent the Liouvillian $\mathcal{L}_n$ and the update map $\mathcal{U}_n$, respectively. Note that we assume that the Liouvillian is time-independent during an update step, which is a valid assumption in the context of numerical methods for the Maxwell-Liouville-von Neumann equations, where the electric field (that causes the time-dependency) and the density matrix are updated alternately.

This update map is completely positive and trace preserving by definition, since the differential equation is solved analytically at every time step and the Liouvillian was chosen to yield a CPTP map as solution~\cite{lindblad1976, gorini1976}. Here, we assume that there are numerical methods that solve the matrix exponential with machine precision so that the CPTP conditions will not be altered by the implementation. Obviously, such methods exist and are implemented in publicly available libraries such as the Eigen library~\cite{eigen-project}. Here, the scaling and squaring method combined with the Pad\'e approximation~\cite{moler2003} is implemented which calculates the exponential of a $M \times M$ matrix in $\mathcal{O}(M^3)$ time. Since in Liouville space the matrices $L_n$ are $N^2 \times N^2$, this approach has the complexity $\mathcal{O}(N^6)$ which is of course a significant drawback. Therefore, the algorithm presented in~\cite{almohy2011} is a promising alternative. It exploits the fact that not the matrix exponential itself but its action on a vector is asked. This action can be determined in $\mathcal{O}(N^4)$ time and up to a user-defined accuracy. The algorithm bases on Krylov subspace methods~\cite{hochbruck1997, hochbruck1998} like the approaches presented in~\cite{pototschnig2009, guduff2017}. A similar technique is described in~\cite{berman1992, kosloff1994, basilewitsch2018}, where the matrix exponential function is expressed using Chebychev polynomials. It should be noted that while the algorithm using the action of the matrix exponential is designed for the application in Liouville space only, the other methods may be used in both Liouville space and regular representation. Then, the performance of the methods may differ depending on the representation. The reason for this is the dimension of the Liouville space, which represents a large hindrance that may only be overcome if the involved matrices are sparse. However, the full analysis of the numerical performance of the methods is beyond the scope of this paper and will be in the focus of future research.

\subsection{Matrix exponential approaches that use approximations}

The evaluation of the matrix exponential function is costly, in particular when the matrices in Liouville space are concerned. Therefore, related work focused on the solution of the Liouville-von Neumann equation in regular representation. A closed analytic expression such as in Eq.~(\ref{eq:update-me}) cannot be derived, but as remedy the symmetric Strang operator splitting technique~\cite{strang1968} can be invoked~\cite{bidegaray2003, saut2006, marskar2011, bidegaray2001}. This approach splits the exponential $\exp(\mathcal{L}_n\Delta t) \approx \exp(\mathcal{L}_1 \Delta t/2) \exp(\mathcal{L}_{2,n}\Delta t) \exp(\mathcal{L}_1 \Delta t/2)$ into two parts that constitute the solution to the time-independent Liouvillian $\mathcal{L}_1$ and the time-dependent Liouvillian $\mathcal{L}_{2,n}$, where $\mathcal{L}_n = \mathcal{L}_1 + \mathcal{L}_{2,n}$. The exponentials represent the solutions of the individual Liouvillians, which can usually be determined separately. This splitting produces an error of order $\mathcal{O}(\Delta t^2)$ (except in the unlikely case where both parts of the Liouvillian commute). However, if each part of the Liouvillian yields a CPTP map as solution, the combination is again a CPTP map.

The separation of the time-dependent part has the advantage -- apart from allowing analytic solutions -- that the time-independent solution has to be determined only once and can be precalculated.
Hence, related literature focused on the efficient evaluation of the time-dependent solution $\exp(\mathcal{L}_{2,n}\Delta t)\hat \rho = \exp(\mathrm{i}\hbar^{-1} \hat V_n \Delta t)\hat \rho\exp(-\mathrm{i}\hbar^{-1} \hat V_n \Delta t)$, where the interaction term $\hat V_n = \hat \mu E_{n+1/2}$ is the product of the dipole moment operator $\hat \mu$ and the (obviously time-dependent) electric field $E_{n+1/2}$, as this evaluation must be performed at every time step.

Since the exponential of an $N \times N$ matrix $\hat V$ can be achieved with a complexity of $\mathcal{O}(N^3)$~\cite{moler2003} and the multiplication of $N \times N$ matrices completes in $\mathcal{O}(N^{\approx 2.37})$ time~\cite{coppersmith1990}, the exponential calculation dominates the complexity of the update step. The goal of the approaches outlined below is to approximate the exponential using matrix multiplications and sums. Then, the complexity is reduced and sparse algorithms may be used (note that the exponential of a matrix is usually dense, even when the matrix itself is sparse).

In~\cite{bidegaray2003, bidegaray2001} the exponential $\exp(\mathrm{i}\hbar^{-1} \hat V_n \Delta t) = \hat A$ is approximated using the Crank-Nicolson scheme. We can readily see that this exponential is the solution of the simple differential equation $\partial_t \hat A = \mathrm{i} \hbar^{-1} \hat V \hat A$. The CN scheme is applied to this differential equation (in contrast to the work in~\cite{ziolkowski1995}, where the CN scheme is applied to the Liouville-von Neumann equation) and the approximation $\exp(\mathrm{i} \hbar^{-1} \hat V \Delta t) \approx (\hat I - \mathrm{i} \hbar^{-1} \hat V \Delta t/2)^{-1}(\hat I + \mathrm{i} \hbar^{-1} \hat V \Delta t/2)$ is derived. This approximation can be related to the Cayley transform of the skew-Hermitian matrix $\mathrm{i}\hat V$, which is guaranteed to be unitary~\cite{courant2004}. Of course, an additional numerical error is introduced, but the density matrix properties are preserved thanks to the clever choice of the approximation. In terms of complexity, the matrix inverse operation and the matrix multiplication are equal~\cite{cormen2009} and the complexity of the update step is $\mathcal{O}(N^{\approx 2.37})$.

In~\cite{songolo2018}, a variation of this approach was described in the context of nonstandard finite difference methods, where the resulting schemes are currently limited to the elementary but essential case with two energy levels. In this case, analytic solutions exist for the general matrix exponential and further simplifications may be applied -- for example by assuming that the main diagonal entries of the dipole moment operator $\hat \mu$ are zero. Here, the aim of any numerical method implementation is to exploit such special cases without loss of generalization of the original problem.

Other approaches (e.g.,~\cite{wu2013ultrafast}) use the Taylor series to evaluate the matrix exponential. According to~\cite{moler2003}, this method converges slowly and will therefore show inferior performance or relatively large numerical errors. As an alternative, one could think of an approximation $\exp(\mathrm{i} \hbar^{-1} \hat V \Delta t) \approx \hat I + \mathrm{i} \hbar^{-1} \hat V \Delta t - (\hbar^{-1} \hat V \Delta t)^2/2 + \dots + (\mathrm{i} \hbar^{-1} \hat V \Delta t)^k/k! = \hat Y$ based on the truncated Taylor series. This way the update step has the form $\hat Y \hat \rho_n \hat Y^\dagger$ but the condition in Eq.~(\ref{eq:cptp2}) is not fulfilled since $\hat Y \hat Y^\dagger = \hat I + \mathcal{O}(\Delta t^l)$, where $l = 2\lfloor 1 + k/2 \rfloor$. Hence, this technique does not feature a CPTP update step.

\subsection{Runge-Kutta method}

Several research groups~\cite{sukharev2011, deinega2014, cartar2017} used the fourth-order Runge-Kutta method (see e.g.,~\cite{hairer1993}) to solve the Liouville-von Neumann equation. Here, the update step reads
\begin{equation}
  \hat \rho_{n+1} = \hat \rho_{n} + \Delta t \left( k_1 + 2 k_2 + 2 k_3 + k_4 \right)/6,
\end{equation}
where $k_1 = \mathcal{L}_n(\hat \rho_{n})$, $k_2 =  \mathcal{L}_{n+1/2}(\hat \rho_{n} + \Delta t k_1/2)$, $k_3 =  \mathcal{L}_{n+1/2}(\hat \rho_{n} + \Delta t k_2/2)$, and $k_4 = \mathcal{L}_{n+1}(\hat \rho_{n} + \Delta t k_3)$. This method is promising since the computational workload of the update step is dominated by multiplications of $N \times N$ matrices (assuming that we apply the Runge-Kutta method in regular representation). Similar to the matrix exponential methods using approximations above, the computational complexity is $\mathcal{O}(N^{\approx 2.37})$ and sparse methods can be applied.

We apply this method to a simple test system with the Liouvillian $\mathcal{L}(\hat \rho) = -\mathrm{i}\hslash^{-1} [ \hat H, \hat \rho ]$, where the Hamiltonian $\hat H$ is time-independent, and transform the update step to the Liouville space. In this representation, the vector $\vec\rho$ is updated using
\begin{equation}
  \vec\rho_{n+1} = \left[ I + L \Delta t + \frac{1}{2} \left(L \Delta t\right)^2 + \frac{1}{6} \left(L \Delta t\right)^3 + \frac{1}{24} \left(L \Delta t\right)^4 \right] \vec\rho_n \coloneqq U \vec\rho_n,
\end{equation}
where $I$ is the $N^2 \times N^2$ identity matrix and the Liouvillian $L = \mathrm{i}\hslash^{-1} ( \hat H^* \otimes \hat I - \hat I \otimes \hat H)$. Here, $\hat I$ is the $N \times N$ identity matrix, $\otimes$ is the Kronecker product, and the asterisk denotes the complex conjugate. See~\cite{marskar2011} for a detailed description of the Liouvillian in Liouville space.

Now we rewrite the update matrix $U$ using $\tilde H = -\mathrm{i}\hslash^{-1}\Delta t \hat H$ and assess whether it can be decomposed into
\begin{equation}
  U = \sum_{j=0}^\infty c_j \sum_{k=0}^j \frac{j!}{k!\left(j - k\right)!} \left( \tilde H^*\right)^{j-k} \otimes \tilde H^{k} \overset{!}{=} \sum_{i=1}^{N^2} \hat V_i^* \otimes \hat V_i,
  \label{eq:update-decomp}
\end{equation}
which is the condition in Eq.~(\ref{eq:cptp}) in Liouville space~\cite{havel2003}. For the Runge-Kutta method, the coefficients $c_j$ on the left-hand side are zero for $j > 4$. Also, it becomes apparent that if such matrices $\hat V_i$ exist, they must be functions of the Hamiltonian $\tilde H$. We assume that $\Delta t$ is chosen sufficiently small so that we can expand each matrix $\hat V_i = \sum_{l=0}^\infty a_{i, l} \tilde H^l$ as Taylor series. Then, the decomposition reads
\begin{equation}
\sum_{i=1}^{N^2} \hat V_i^* \otimes \hat V_i = \sum_{i=1}^{N^2} \sum_{l=0}^\infty \sum_{k=0}^\infty a_{i, l}^* \left(\tilde H^*\right)^l \otimes a_{i, k} \tilde H^k = \sum_{l=0}^\infty \sum_{k=0}^\infty \sum_{i=1}^{N^2} a_{i, l}^* a_{i, k} \left(\tilde H^*\right)^l \otimes  \tilde H^k.
\label{eq:decomp}
\end{equation}
We note that since $c_6 = 0$ for the Runge-Kutta method, for $j = 6$ and $k = 3$ the corresponding term in Eq.~(\ref{eq:update-decomp}) vanishes, and consequently the term with the powers $l = k = 3$ in Eq.~(\ref{eq:decomp}) should also vanish. Therefore, the sum $\sum_{i=1}^{N^2} | a_{i, 3} |^2 = 0$ and subsequently all coefficients $a_{i, 3}$ must be zero. However, a term with the powers $j = 3$ and $k = 0$ is present in the update matrix in Eq.~(\ref{eq:update-decomp}), but the corresponding term (with $l = 3$ and $k = 0$) in Eq.~(\ref{eq:decomp}) vanishes if all $a_{i, 3}$ are zero. We deduce that the update matrix cannot be decomposed and the update map of the Runge-Kutta method is not CPTP.

\subsection{Predictor-corrector technique}

In their work, Ziolkowski, Slavcheva et al.~\cite{ziolkowski1995, slavcheva2002} treat the Liouville-von Neumann equation with the Crank-Nicolson scheme. The positivity preservation of this scheme has been discussed in~\cite{bidegaray2001}. However, the actual implementation uses the predictor-corrector technique to resolve the implicit nature of Crank-Nicolson. Hence, we shall concentrate on the explicit predictor-corrector method in this work.

The predictor-corrector update step begins by setting $\vec\rho_\mathrm{PC} = \vec\rho_n$ and then executes the procedure
\begin{equation}
  \vec \rho_\mathrm{PC} \leftarrow \vec \rho_n + \frac{\Delta t}{2} L \left( \vec\rho_\mathrm{PC} + \vec \rho_n \right)
\end{equation}
four times. Then, the result is assigned to the value $\vec \rho_{n+1} = \vec \rho_\mathrm{PC}$. Again, we consider a simple test system with a time-independent Liouvillian that only contains the commutator part and write for the complete update step in Liouville space
\begin{equation}
   \vec\rho_{n+1} = \left[ I + L \Delta t + \frac{1}{2} \left(L \Delta t\right)^2 + \frac{1}{4} \left(L \Delta t\right)^3 + \frac{1}{8} \left(L \Delta t\right)^4 \right] \vec\rho_n \coloneqq U \vec\rho_n.
\end{equation}
We can readily see that apart from different coefficients $c_j$ the predictor-corrector technique and the Runge-Kutta method have the same update step. Therefore, we can deduce that the computational complexity is the same (the predictor-corrector method can be implemented in regular representation as well and sparse methods can be applied) and the update map of the predictor-corrector technique is not CPTP. Indeed, by using the argumentation above, one can show that no method of the form
\begin{equation}
  \vec\rho_{n+1} = \sum_{j=0}^M c_j \left(L \Delta t \right)^j \vec\rho_n
\end{equation}
with a finite number of steps $M$ can be decomposed to fulfill the condition in Eq.~(\ref{eq:cptp}).

\section{Verification}

As we have shown in the section before, only one of the typically used numerical methods -- namely, the matrix exponential method -- can be represented as completely positive trace preserving map and is therefore guaranteed to yield realistic results in long-term simulations. However, it remains to be demonstrated that the results of this theoretical analysis are relevant in a practical example. In this section, we implement a simple simulation based on related literature and compare the results of the different numerical methods.

Similar to our simple test system before, we only consider a time-independent Hamiltonian
\begin{equation}
\hat H = \begin{bmatrix}
  0 & \mu E & 0 & 0 & 0 & 0\\
  \mu E & \hslash \omega_{12} & \mu E & 0 & 0 & 0\\
  0 & \mu E & H_{22} + \hslash \omega_{23} & \mu E & 0 & 0\\
  0 & 0 & \mu E & H_{33} + \hslash \omega_{34} & \mu E & 0\\
  0 & 0 & 0 & \mu E & H_{44} + \hslash \omega_{45} & \mu E\\
  0 & 0 & 0 & 0 & \mu E & H_{55} + \hslash \omega_{56}
\end{bmatrix},
\end{equation}
which describes a system with six energy levels $H_{ii}$ separated by $\hslash \omega_{i, i+1} = \hslash \omega_0 [1 - 0.1(i - 3)]$, where $\omega_0 = 2 \pi \times \SI{e13}{\per\second}$. The influence of the constant electric field $E = \SI{9e9}{\volt\per\meter}$ on the system is modeled using the dipole approximation and the dipole moment $\mu = \SI{e-29}{\ampere\second\meter}$. This setup corresponds to the six-level anharmonic ladder setup in~\cite{marskar2011}. We selected this particular example since it is a well-established, multi-level setup that constitutes a significant challenge for the numerical methods in terms of accuracy and performance.

The three different numerical methods were implemented in MATLAB. By using its variable-precision arithmetic (vpa) toolbox, we increased the precision and could assure that the results below are not affected by round-off error artifacts. All methods used the same time step size $\Delta t = \SI{0.1}{\femto\second}$, which was chosen sufficiently small in order to avoid stability issues. The resulting MATLAB scripts are publicly available as open-source project~\cite{riesch2019positivitycode}.

\begin{figure}
\centering
\textbf{a)}\includegraphics[width=0.45\textwidth]{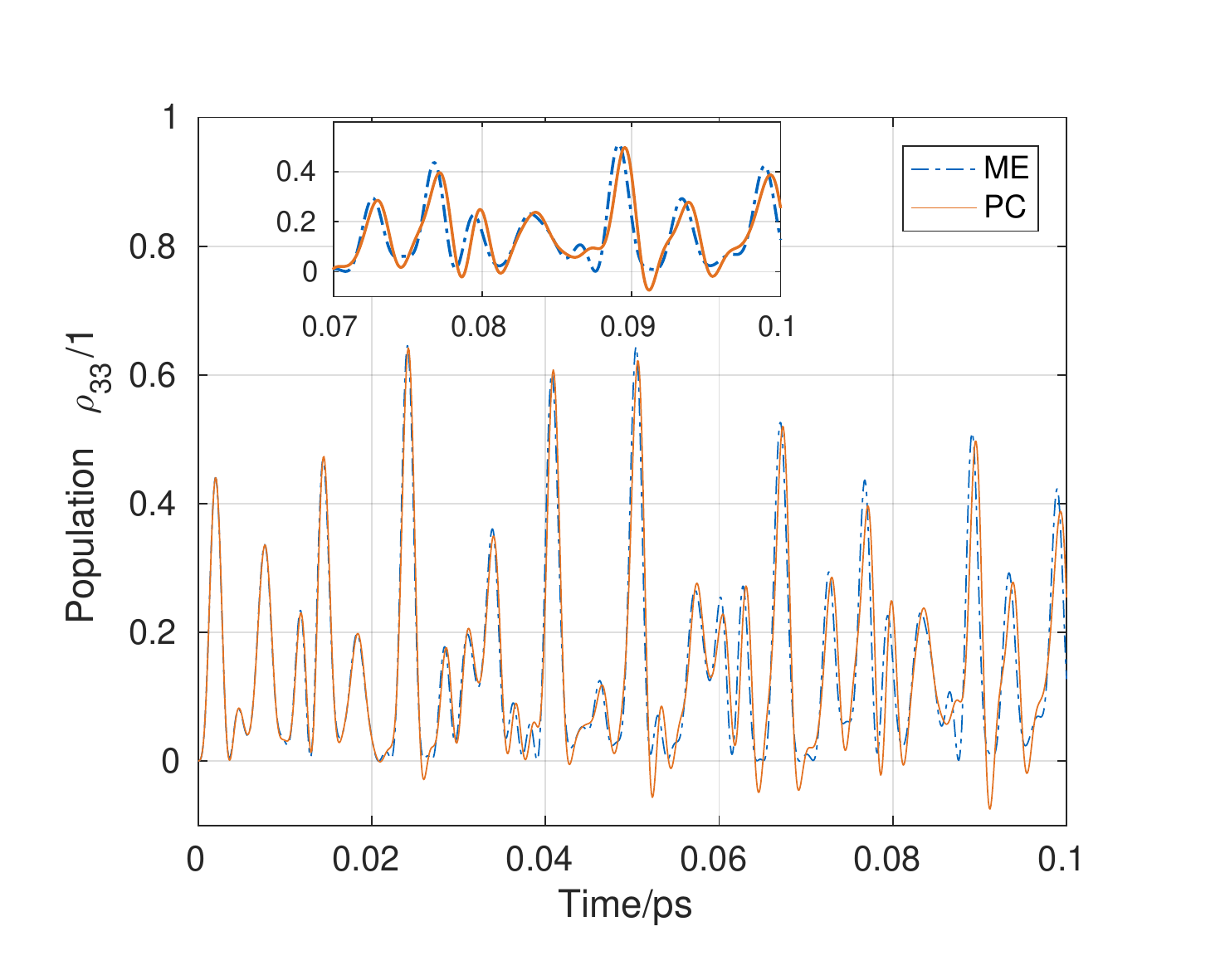}~
\textbf{b)}\includegraphics[width=0.45\textwidth]{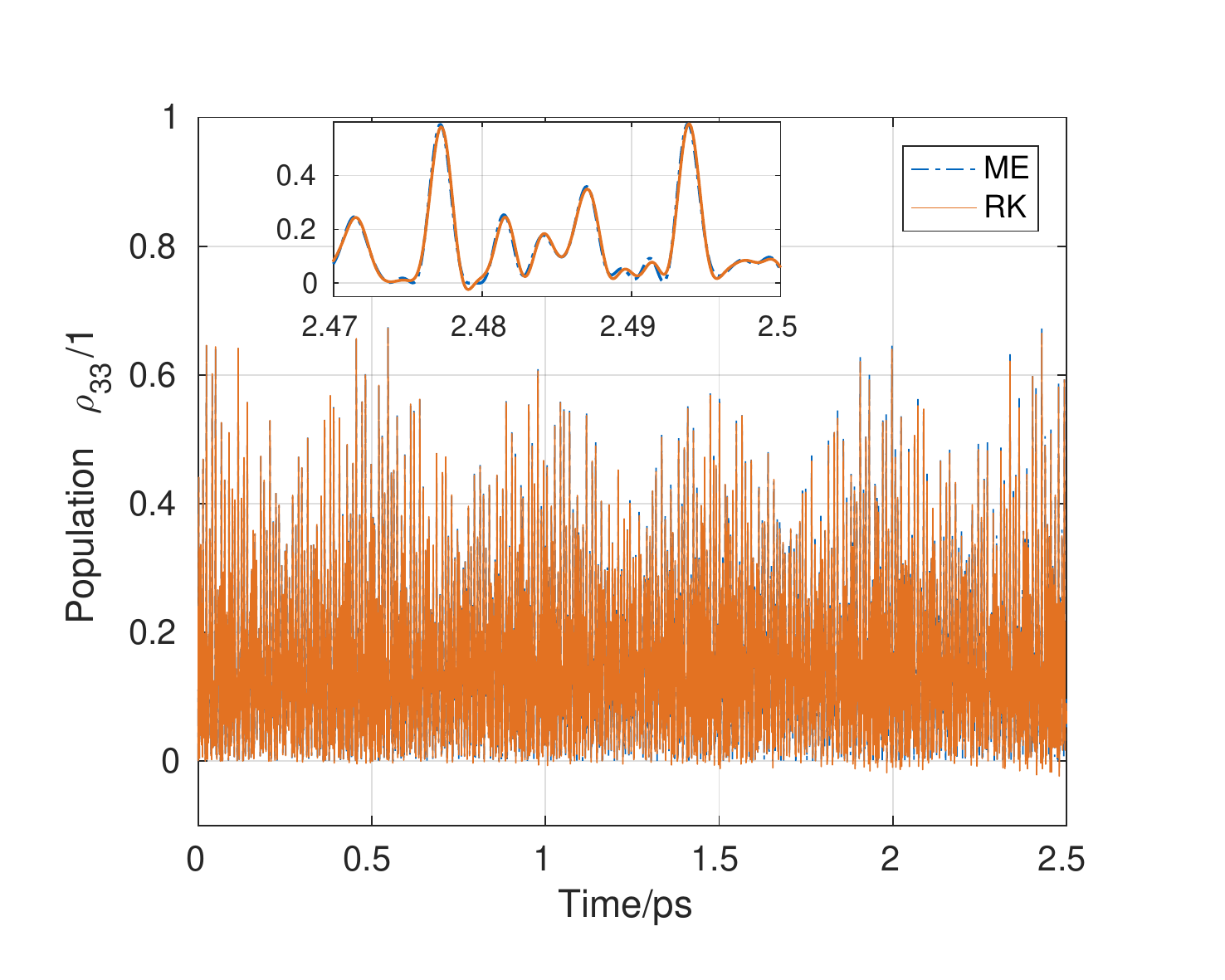}
\caption{Simulation results of the anharmonic ladder system using \textbf{a)} the predictor-corrector approach and \textbf{b)} the Runge-Kutta method. The results of the matrix exponential method serve as reference in both parts of the figure. For details of the curves see the respective inset.}
\label{fig:results}
\end{figure}

Figure~\ref{fig:results} depicts the simulation results of all three methods. The matrix exponential method solves the problem exactly and serves as reference. By close inspection we can see that the population $\rho_{33}$ remains in the interval $[0, 1]$, as the theory dictates. In contrast, the results of the predictor-corrector method (Fig.~\ref{fig:results}a) show that after a short duration the population becomes negative, which is clearly a violation of the properties of the density matrix. Similarly, the population becomes negative when using the Runge-Kutta method (see Fig.~\ref{fig:results}b). In this case, the first instance of a negative population occurs at a later point in time, which is consistent with the order of accuracy of the methods (a fourth-order Runge-Kutta method was used, the predictor-corrector approach is a second-order method). Nevertheless, the Runge-Kutta method may yield unrealistic results for certain simulation end times.

Finally, it should be noted that although the population $\rho_{33}$ was chosen as figure of merit, the discussed features are also visible in other populations.

\section{Conclusion}

The density matrix is updated from time step to time step when solving the (Maxwell-)Liouville-von Neumann equation(s). At all time steps, the properties of the density matrix must be preserved. This is guaranteed if the update step is a completely positive trace preserving (CPTP) map. In the work at hand, we established the criterion that only numerical methods that feature a CPTP update map can be considered for future use in long-term simulations.

Subsequently, we analyzed the three groups of numerical methods that are dominant in related literature and found that only the group of matrix exponential (ME) methods fulfill this criterion. The results of the theoretical analysis were later confirmed by simulating a well-established test setup from literature. Therefore, we shall focus on this group of methods in future.

However, the evaluation of the matrix exponential is still costly, in particular for a large number of energy levels. Hence, future research will focus on the further optimization of the matrix exponential calculation as well as the assessment of alternative numerical methods with respect to the preservation of the properties of the density matrix.

\section*{Acknowledgments}

This work was supported by the German Research Foundation (DFG) within the Heisenberg program (JI 115/4-2) and under DFG Grant No. JI 115/9-1. The authors thank Michael Haider for the stimulating discussions and his input on the CPTP maps as well as Christiane Koch for pointing out the Chebychev polynomial methods.

\bibliographystyle{osajnl}
\bibliography{literature}

\begin{thebibliography}{10}
\newcommand{\enquote}[1]{``#1''}

\bibitem{tang2005}
C.~L. Tang, \emph{Fundamentals of Quantum Mechanics: For Solid State
  Electronics and Optics} (Cambridge University Press, 2005).

\bibitem{jirauschek2014}
C.~Jirauschek and T.~Kubis, \enquote{Modeling techniques for quantum cascade
  lasers,} Appl. Phys. Rev. \textbf{1}, 011307 (2014).

\bibitem{riesch2018atrasc1}
M.~Riesch, P.~Tzenov, and C.~Jirauschek, \enquote{Dynamic simulations of
  quantum cascade lasers beyond the rotating wave approximation,} in
  \enquote{Proceedings of 2nd URSI AT-RASC,}  (2018).

\bibitem{ziolkowski1995}
R.~W. Ziolkowski, J.~M. Arnold, and D.~M. Gogny, \enquote{Ultrafast pulse
  interactions with two-level atoms,} Phys. Rev. A \textbf{52}, 3082--3094
  (1995).

\bibitem{slavcheva2002}
G.~Slavcheva, J.~M. Arnold, I.~Wallace, and R.~W. Ziolkowski, \enquote{Coupled
  {Maxwell-pseudospin} equations for investigation of self-induced transparency
  effects in a degenerate three-level quantum system in two dimensions:
  {Finite-difference} time-domain study,} Phys. Rev. A \textbf{66}, 63418
  (2002).

\bibitem{hellsing1986}
B.~Hellsing and H.~Metiu, \enquote{An efficient method for solving the quantum
  {Liouville} equation: Applications to electronic absorption spectroscopy,}
  Chem. Phys. Lett. \textbf{127}, 45--49 (1986).

\bibitem{berman1992}
M.~Berman, R.~Kosloff, and H.~Tal-Ezer, \enquote{Solution of the time-dependent
  {Liouville-von Neumann} equation: dissipative evolution,} J. Phys. A: Math.
  Gen. \textbf{25}, 1283--1307 (1992).

\bibitem{bidegaray2003}
B.~Bid{\'e}garay, \enquote{Time discretizations for {Maxwell-Bloch} equations,}
  Numer. Methods Partial Differ. Equ. \textbf{19}, 284--300 (2003).

\bibitem{saut2006}
O.~Saut and A.~Bourgeade, \enquote{Numerical methods for the bidimensional
  {Maxwell--Bloch} equations in nonlinear crystals,} J. Comput. Phys.
  \textbf{213}, 823--843 (2006).

\bibitem{marskar2011}
R.~Marskar and U.~{\"O}sterberg, \enquote{Multilevel {Maxwell-Bloch}
  simulations in inhomogeneously broadened media,} Opt. Express \textbf{19},
  16784--16796 (2011).

\bibitem{garraway1994}
B.~Garraway and P.~Knight, \enquote{Comparison of quantum-state diffusion and
  quantum-jump simulations of two-photon processes in a dissipative
  environment,} Phys. Rev. A \textbf{49}, 1266 (1994).

\bibitem{sukharev2011}
M.~Sukharev and A.~Nitzan, \enquote{Numerical studies of the interaction of an
  atomic sample with the electromagnetic field in two dimensions,} Phys. Rev. A
  \textbf{84}, 043802 (2011).

\bibitem{deinega2014}
A.~Deinega and T.~Seideman, \enquote{Self-interaction-free approaches for
  self-consistent solution of the {Maxwell-Liouville} equations,} Phys. Rev. A
  \textbf{89}, 022501 (2014).

\bibitem{cartar2017}
W.~Cartar, J.~M\o{}rk, and S.~Hughes, \enquote{Self-consistent {Maxwell-Bloch}
  model of quantum-dot photonic-crystal-cavity lasers,} Phys. Rev. A
  \textbf{96}, 023859 (2017).

\bibitem{kraus1983}
K.~Kraus, A.~B{\"o}hm, J.~Dollard, and W.~Wootters, eds., \emph{States,
  Effects, and Operations: {Fundamental} Notions of Quantum Theory. {Lectures}
  in Mathematical Physics at the {University of Texas at Austin}}, Lecture
  Notes in Physics (Springer, Berlin, Heidelberg, 1983).

\bibitem{bidegaray2001}
B.~Bid{\'e}garay, A.~Bourgeade, and D.~Reignier, \enquote{Introducing physical
  relaxation terms in {Bloch} equations,} J. Comput. Phys. \textbf{170},
  603--613 (2001).

\bibitem{songolo2018}
M.~E. Songolo and B.~Bid{\'e}garay-Fesquet, \enquote{Nonstandard
  finite-difference schemes for the two-level {Bloch} model,} Int. J. Model.
  Simul. Sci. Comput. p. 1850033 (2018).

\bibitem{riesch2018oqel}
M.~Riesch, N.~Tchipev, S.~Senninger, H.-J. Bungartz, and C.~Jirauschek,
  \enquote{Performance evaluation of numerical methods for the
  {Maxwell--Liouville--von Neumann} equations,} Opt. Quant. Electron.
  \textbf{50}, 112 (2018).

\bibitem{lindblad1976}
G.~Lindblad, \enquote{On the generators of quantum dynamical semigroups,}
  Commun. Math. Phys. \textbf{48}, 119--130 (1976).

\bibitem{gorini1976}
V.~Gorini, A.~Kossakowski, and E.~C.~G. Sudarshan, \enquote{Completely positive
  dynamical semigroups of {$N$}-level systems,} J. Math. Phys. \textbf{17},
  821--825 (1976).

\bibitem{lebellac2011}
M.~Le~Bellac, \emph{Quantum Physics} (Cambridge University Press, 2011).

\bibitem{kraus1971}
K.~Kraus, \enquote{General state changes in quantum theory,} Ann. Phys.
  \textbf{64}, 311--335 (1971).

\bibitem{choi1975}
M.-D. Choi, \enquote{Completely positive linear maps on complex matrices,}
  Linear Algebra Its Appl. \textbf{10}, 285--290 (1975).

\bibitem{nielsen2010}
M.~A. Nielsen and I.~L. Chuang, \emph{Quantum Computation and Quantum
  Information} (Cambridge University Press, 2010), 10th ed.

\bibitem{eigen-project}
G.~Guennebaud, B.~Jacob \emph{et~al.}, \enquote{Eigen v3,}
  http://eigen.tuxfamily.org (2010).

\bibitem{moler2003}
C.~Moler and C.~V. Loan, \enquote{Nineteen dubious ways to compute the
  exponential of a matrix, twenty-five years later,} SIAM Rev. \textbf{45},
  3--49 (2003).

\bibitem{almohy2011}
A.~H. Al-Mohy and N.~J. Higham, \enquote{Computing the action of the matrix
  exponential, with an application to exponential integrators,} SIAM J. Sci.
  Comput. \textbf{33}, 488--511 (2011).

\bibitem{hochbruck1997}
M.~Hochbruck and C.~Lubich, \enquote{On {Krylov} subspace approximations to the
  matrix exponential operator,} SIAM J. Numer. Anal. \textbf{34}, 1911--1925
  (1997).

\bibitem{hochbruck1998}
M.~Hochbruck, C.~Lubich, and H.~Selhofer, \enquote{Exponential integrators for
  large systems of differential equations,} SIAM J. Sci. Comput. \textbf{19},
  1552--1574 (1998).

\bibitem{pototschnig2009}
M.~Pototschnig, J.~Niegemann, L.~Tkeshelashvili, and K.~Busch,
  \enquote{Time-domain simulations of the nonlinear {Maxwell} equations using
  operator-exponential methods,} IEEE Trans. Antennas Propag. \textbf{57},
  475--483 (2009).

\bibitem{guduff2017}
L.~Guduff, A.~J. Allami, C.~van Heijenoort, J.-N. Dumez, and I.~Kuprov,
  \enquote{Efficient simulation of ultrafast magnetic resonance experiments,}
  Phys. Chem. Chem. Phys. \textbf{19}, 17577--17586 (2017).

\bibitem{kosloff1994}
R.~Kosloff, \enquote{Propagation methods for quantum molecular dynamics,} Annu.
  Rev. Phys. Chem. \textbf{45}, 145--178 (1994).

\bibitem{basilewitsch2018}
D.~Basilewitsch, L.~Marder, and C.~P. Koch, \enquote{Dissipative quantum
  dynamics and optimal control using iterative time ordering: an application to
  superconducting qubits,} Eur. Phys. J. B \textbf{91}, 161 (2018).

\bibitem{strang1968}
G.~Strang, \enquote{On the construction and comparison of difference schemes,}
  SIAM J. Numer. Anal. \textbf{5}, 506--517 (1968).

\bibitem{coppersmith1990}
D.~Coppersmith and S.~Winograd, \enquote{Matrix multiplication via arithmetic
  progressions,} J. Symb. Comput. \textbf{9}, 251--280 (1990).

\bibitem{courant2004}
R.~Courant and D.~Hilbert, \emph{Methods of Mathematical Physics}, vol.~1
  (Wiley-VCH, 2004).

\bibitem{cormen2009}
T.~H. Cormen, C.~E. Leiserson, R.~L. Rivest, and C.~Stein, \emph{Introduction
  to Algorithms} (The MIT Press, 2009), 3rd ed.

\bibitem{wu2013ultrafast}
M.~Wu, S.~Chen, K.~J. Schafer, and M.~B. Gaarde, \enquote{Ultrafast
  time-dependent absorption in a macroscopic three-level helium gas,} Phys.
  Rev. A \textbf{87}, 013828 (2013).

\bibitem{hairer1993}
E.~Hairer, S.~P. N\o{}rsett, and G.~Wanner, \emph{Solving Ordinary Differential
  Equations I} (Springer-Verlag Berlin Heidelberg, 1993), 2nd ed.

\bibitem{havel2003}
T.~F. Havel, \enquote{Robust procedures for converting among {Lindblad},
  {Kraus} and matrix representations of quantum dynamical semigroups,} J. Math.
  Phys. \textbf{44}, 534--557 (2003).

\bibitem{riesch2019positivitycode}
M.~Riesch, \enquote{Supplementary {MATLAB} code,}  (2019).
  \url{https://doi.org/10.5281/zenodo.2560305}.

\end{thebibliography}

\end{document}